\documentclass[acmsmall, screen]{acmart}

\usepackage{tabularx,arydshln}
\usepackage{multirow}

\usepackage{xargs} 
\usepackage{soul}  
\usepackage{color} 
\usepackage{xspace}
\usepackage{listings}

\newcommand{\eg}{{\it e.g.,\ }}
\newcommand{\etal}{{\it et al.\ }}

\newcommand{\ie}{{\it i.e.,\ }}

\newcommand{\aka}{{a.k.a.}\xspace}
\newcommand{\ipstart}[1]{\vspace{1mm} \noindent{\textit{#1.}}}

\definecolor{activegold}{RGB}{255,193,61}

\definecolor{lightorange}{RGB}{230, 170, 50}
\definecolor{lightgreen}{RGB}{121,210,121}
\definecolor{lightteal}{RGB}{121,199,210}
\definecolor{lightblue}{RGB}{100,212,239}
\definecolor{lightpurple}{RGB}{153,102,255}
\definecolor{lightred}{RGB}{245, 132, 120}
\definecolor{red}{RGB}{178,34,34}
\definecolor{gray}{RGB}{166,166,166}

\newcommandx{\guest}[3][1=]
    {\setulcolor{lightorange}{\ul{#1}} \textcolor{lightorange}
    {[\textbf{#2:} #3]}}
\newcommandx{\taewook}[2][1=]
    {\setulcolor{lightgreen}{\ul{#1}} \textcolor{lightgreen}
    {[\textbf{Taewook:} #2]}}
\newcommandx{\matt}[2][1=]
    {\setulcolor{lightred}{\ul{#1}} \textcolor{lightred}
    {[\textbf{Matt:} #2]}}
\newcommandx{\maia}[2][1=]
    {\setulcolor{lightteal}{\ul{#1}} \textcolor{lightteal}
    {[\textbf{Maia:} #2]}}
\newcommandx{\angela}[2][1=]
    {\setulcolor{lightpurple}{\ul{#1}} \textcolor{lightpurple}
    {[\textbf{Angela:} #2]}}
\newcommandx{\hyeok}[2][1=]
    {\setulcolor{lightblue}{\ul{#1}} \textcolor{lightblue}
    {[\textbf{Hyeok:} #2]}}

\definecolor{forestgreen(web)}{rgb}{0.13, 0.55, 0.13}
\definecolor{freespeechblue}{rgb}{0.2852, 0.3789, 0.8008}
\definecolor{slateblue}{rgb}{0.4258, 0.2969, 0.7734}
\definecolor{mahogany}{rgb}{0.8359, 0.1719, 0.1758}
\definecolor{deep}{rgb}{0.8789, 0.2188, 0.4688}
\definecolor{Issue1}{HTML}{ff7e16}
\definecolor{Issue2}{HTML}{4860cc}
\definecolor{Issue3}{HTML}{6c4bc5}
\definecolor{Issue4}{HTML}{d52b2c}
\definecolor{Issue4a}{HTML}{BF9000}
\definecolor{Issue5}{HTML}{327FCC}
\definecolor{Issue6}{HTML}{E69138}
\definecolor{Issue7}{HTML}{e03777}
\definecolor{Issue8}{HTML}{148f6a}
\definecolor{Issue9}{HTML}{bbbc27}
\definecolor{Issue10}{HTML}{dc4676}

\newif\iftrack
\tracktrue 
\newcommandx{\revised}[1][]{\iftrack{\textcolor{black}{#1}}\else#1\fi}

\AtBeginDocument{%
  \providecommand\BibTeX{{%
    \normalfont B\kern-0.5em{\scshape i\kern-0.25em b}\kern-0.8em\TeX}}}

\begin{document}


\title{Opportunities in Mental Health Support for Informal Dementia Caregivers Suffering from Verbal Agitation}

%


\author{Taewook Kim}
\email{taewook@u.northwestern.edu}
\affiliation{%
 \institution{Northwestern University}
 \city{Evanston}
 \state{IL}
 \country{USA}
}

\author{Hyeok Kim}
\email{hyeokkim2024@u.northwestern.edu}
\affiliation{%
 \institution{Northwestern University}
 \city{Evanston}
 \state{IL}
 \country{USA}
}

\author{Angela Roberts}
\email{angela.roberts@uwo.ca}
\affiliation{%
 \institution{University of Western Ontario}
 \city{London}
 \state{Ontario}
 \country{Canada}
}

\author{Maia Jacobs}
\email{maia.jacobs@northwestern.edu}
\affiliation{%
 \institution{Northwestern University}
 \city{Evanston}
 \state{IL}
 \country{USA}
}

\author{Matthew Kay}
\email{mjskay@northwestern.edu}
\affiliation{%
 \institution{Northwestern University}
 \city{Evanston}
 \state{IL}
 \country{USA}
}


\begin{abstract}
People with dementia (PwD) often present verbal agitation such as cursing, screaming, and persistently complaining. Verbal agitation can impose mental distress on informal caregivers (\eg~family, friends), which may cause severe mental illnesses, such as depression and anxiety disorders. To improve informal caregivers' mental health, we explore design opportunities by interviewing 11 informal caregivers suffering from verbal agitation of PwD. In particular, we first characterize how the predictability of verbal agitation impacts informal caregivers' mental health and how caregivers' coping strategies vary before, during, and after verbal agitation. Based on our findings, we propose design opportunities to improve the mental health of informal caregivers suffering from verbal agitation: distracting PwD (in-situ support; before), prompting just-in-time maneuvers (information support; during), and comfort and education (social \& information support; after). We discuss our reflections on cultural disparities between participants. Our work envisions a broader design space for supporting informal caregivers' well-being and describes when and how that support could be provided.
\end{abstract}




\maketitle

\section{introduction}

People with dementia (PwD) often exhibit \textit{verbal agitation}, a major behavioral symptom of dementia that includes screaming, swearing, and repeating phrases~\cite{cohen-mansfield2001:intervention,sefcik2019:review,bedard2011:agitation}. Verbal agitation is more common for people in advanced stages of Alzheimer's and vascular dementia~\cite{hurley1992:discomfort,scarmeas2007:predictor,sefcik2020:vocalization,suh2004:validation}, in conjunction with other psychological symptoms such as aggression and disinhibition~\cite{sefcik2020:vocalization,selbaek2013:neuropsychiatric}. Such verbal agitation often causes significant mental distress in caregivers~\cite{brodaty1990:carers,haley1987:consequences}, which may impede them from performing their essential care activities (\eg~changing a diaper, assisting with eating).

Informal caregivers, such as family members, spouses, or friends, are a primary care source for many PwD~\cite{quinn2010:review,devugt2013:early,etters2008:burden}. They often suffer from mental distress due to verbal agitation, impeding their care responsibilities for PwD~\cite{draper2000:casecontrol,hansen2018:reaction,savundranayagam2005:communication}. For instance, informal caregivers often experience cursing or threats from PwD while providing daily personal care, such as changing diapers and brushing teeth~\cite{desai2001:recognition,schreiner2001:agressive}. A substantial number of informal caregivers experience serious mental health issues such as depression~\cite{cohen1988:depression} and anxiety disorder~\cite{dura1991:anxiety, hwang2021impact}. The poor mental health conditions of informal caregivers impacted by verbal agitation can be a barrier to sustainable and comprehensive care for their loved ones with dementia. In addition, many informal caregivers lack access to helpful resources and professional education on dementia care practices, as they are caregivers by necessity, not by training~\cite{prince2013:report,pleasant2020:online}.
Hence, exploring design opportunities for informal caregivers to handle verbal agitation from PwD and manage their mental health is an important step toward developing technological support to better sustain their care activities.

With this in mind, we interviewed 11 informal caregivers who suffer from verbal agitation of PwD in South Korea and the USA. We asked them in what contexts they face verbal agitation of PwD, how they cope with verbal agitation, and what technological support might help improve their mental well-being. We confirmed that verbal agitation has a significant negative impact on the physical and social health of informal caregivers, exacerbating their mental health. We describe how support mechanisms should be tailored to address the different needs of \textit{predictable} and \textit{unpredictable} contexts of verbal agitation from PwD. We identify that caregivers' different coping methods \textit{before} verbal agitation, \textit{during} verbal agitation, and \textit{after} verbal agitation.

Based on these two axes, \textbf{predictability} and \textbf{timing}, we frame design opportunities for systems to support the mental health of informal caregivers suffering from verbal agitation of PwD. In addition, we outline how our design opportunities align with Chen et al.'s~\cite{chen2013:integrality} and Schorch et al.'s~\cite{schorch2016:designing} existing frameworks for providing holistic support for informal caregivers, covering  three types of support: \textit{social support}, \textit{information support}, and \textit{coordination support}. We also add a fourth type of support, \textit{in-situ support}, based on a series of work introducing systems that directly engage the caregiving situation to support dementia caregivers~\cite{guan2021:cue}. From our two axes and these four types of support, we propose the following design opportunities: (1) facilitating informal caregivers' preparation by providing distractions (\textit{in-situ support}) or encouragement (\textit{social \& information support}) \textit{before} \textit{predictable} verbal agitation, (2) bolstering resilience by alleviating impact of verbal agitation (\textit{in-situ support}) or prompting just-in-time maneuvers (\textit{information support}) \textit{during} \textit{unpredictable} verbal agitation, (3) comforting informal caregivers shortly \textit{after} verbal agitation (\textit{social support}) or (4) helping informal caregivers learn and prepare for future events in the distal moment \textit{after} verbal agitation (\textit{information support}).

\section{background}

\subsection{Verbal Agitation of People with Dementia}

Verbal agitation, which includes behaviors such as swearing, threatening, blaming, repeating phrases, and complaining persistently, is a common symptom of dementia~\cite{nagaratnam2003:screaming,talerico2002:aggression,tueth1995:dementia,cariaga1991:vocalization,sefcik2020:vocalization}. This behavioral symptom is particularly prevalent in people with advanced stages of Alzheimer's disease and vascular dementia~\cite{hurley1992:discomfort,scarmeas2007:predictor,sefcik2020:vocalization,suh2004:validation}. PwD often express their psychological, physical, and physiological discomfort through verbal agitation~\cite{kovach1999:assessment}. However, their cognitive impairment, such as memory failure and delusions, can lead them to imagine sources of discomfort that do not actually exist~\cite{tueth1995:dementia}. A typical example can be observed when PwD persistently accuses a family member or often a third person (\eg nurses) of stealing their possessions, even though no items have been taken. This type of paranoia is usually not grounded in reality but originates from delusional beliefs that are a result of their cognitive impairment~\cite{tueth1995:dementia}. Hence, verbal agitation of PwD is considered one of the most challenging behavioral symptoms to manage~\cite{marianne2010:review} as it disturbs caregivers and many other people around them.

\subsection{Impacts of Verbal Agitation on Mental Health of Informal Caregivers of PwD}
\label{rl:impact}

Many PwD receive primary care from informal caregivers who are mostly family, spouses, or friends~\cite{quinn2010:review,devugt2013:early,etters2008:burden}. As a primary care source, informal caregivers suffer from significant mental distress~\cite{brodaty1990:carers,haley1987:consequences}, bearing a huge amount of the physical, socio-psychological, and financial burdens for dementia care~\cite{zarit1986:subjective}. As dementia is progressive and currently incurable, caregivers may face increasing burdens as the disease advances, leading to physical and emotional exhaustion~\cite{conde-sala2014:threeyear, connors2020:burden}.

Verbal agitation aggravates such informal caregivers' mental distress in several aspects. First, it can further complicate daily care activities and exacerbate caregivers' mental distress~\cite{draper2000:casecontrol, hansen2018:reaction, savundranayagam2005:communication}. Dementia caregivers frequently experience verbal agitation of PwD when providing essential personal care, such as toileting, eating, and dressing~\cite{cohen-mansfield1990:screaming, cipriani2011:aggressive, schreiner2001:agressive}. For instance, effective communication between informal caregivers and PwD becomes difficult due to verbal agitation and the cognitive impairment of PwD~\cite{schoenmakers2010:factors,bryan2006:communication}. Such communication barriers can significantly increase the strain on caregivers, exacerbating the difficulties inherent in dementia care~\cite{savundranayagam2005:communication}. Secondly, as dementia is a progressive degeneration of cognitive abilities, informal caregivers may experience more verbal agitation because of increasing types of such caring activities for PwD~\cite{donaldson1997:impact}. Lastly, verbal agitation can strain social relationships between caregivers, the person with dementia, and other family members. For example, PwD may make false accusations of mistreatment or abuse against their informal caregivers, even though no harm has ever been given. However, other people (\eg family members) may not be aware of PwD's repeated paranoid accusations against others~\cite{tueth1995:dementia}. Consequently, informal caregivers may lose their social reputations.

While previous research has identified certain situations (\eg toileting) where verbal agitation may occur, we need a more in-depth understanding of these contexts to design technological support for informal caregivers’ mental health. The specific challenges and stress faced by informal caregivers may vary significantly depending on the nature of the agitation---for example, during routine tasks like changing diapers, compared to dealing with accusations against family members. Given these variances, it is reasonable to suggest that the design for supportive technologies should be tailored to meet these distinct challenges, thereby offering more precise and effective assistance to caregivers. Hence, we decided to investigate the specific situations where informal caregivers encounter verbal agitation, and how they emotionally respond and handle such instances.

\subsection{Supporting Informal Caregivers' Well-being: Social, Information, and Coordination}
\label{rl:framework}

Prior work in CSCW introduces a framework for supporting informal caregivers of people with chronic diseases such as dementia and diabetes. By interviewing informal caregivers, Chen~\etal and Schorch~\etal identify three design considerations---\textit{social support, information support}, and \textit{coordination support}---for improving the well-being of informal caregivers~\cite{chen2013:integrality, schorch2016:designing}.

Having supportive social connections is crucial to the well-being of informal caregivers (\ie~\textbf{social support})~\cite{chen2013:integrality, schorch2016:designing}. Medical research suggests that sharing experiences and feelings with other caregivers in a peer support group is helpful for improving the mental health of informal caregivers~\cite{charlesworth1218:peer}. For example, local libraries provide offline social connections (\eg~a caregiver support community) for informal caregivers~\cite{dai2021:library}. However, such offline social support services are not always available at any time~\cite{zarit1982:families}. Alternatively, informal caregivers can look for social connections from online social media~\cite{johnson2022:lonely}. Prior studies in CSCW show the caregivers' use of social media for support. Johnson~\etal find that people \textit{without} dementia (\ie~family and caregivers) more proactively seek social support than PwD in online forums designated for PwD~\cite{johnson2020:roles}. Lazar~\etal show that the non-judgmental nature of dementia online communities makes people feel comfortable with sharing their experiences there~\cite{lazar2019:safe}.

Informal caregivers often want to better understand their own caregiving experiences and to learn more about their care-receivers and good care practices (\ie \textbf{information support})~\cite{chen2013:integrality, schorch2016:designing}. Counseling and cognitive behavioral therapy (CBT) can help informal caregivers reflect on their own experiences and improve their well-being~\cite{joseph2015:effects, losada2015:cbt}. Caregivers can opt to participate in training programs to be more sensitive to underlying causes of verbal agitation with professional learning sources~\cite{magai2002:impact}. However, to see meaningful outcomes, they have to take part in such training programs for extensive periods of time (\eg~10 one-hour lectures over two weeks~\cite{magai2002:impact}), which may interfere with their day-to-day care activities. Also, informal caregivers often cannot afford to receive such professional support on a regular basis due to the lack of resources such as time and money~\cite{seiji2017:generating}. Instead, informal caregivers may seek information support from online forums~\cite{mentis2020:illusion}.

Informal caregivers need to balance their numerous duties as a caregiver and responsibilities as an individual being, because being a primary caregiver, a parent, and a full-time worker can involve daunting tasks for them (\ie \textbf{coordination support})~\cite{chen2013:integrality, schorch2016:designing}. Prior work in CSCW proposed several ways to alleviate burdens arising from the multiplicity of responsibilities. For example, informal caregivers for the same PwD can utilize a shared-calendar system to coordinate tasks more efficiently~\cite{eschler2015:shared}. Soubutts~\etal show that smart home technologies, such as IoT sensor systems and voice assistants, can facilitate shared caregiving in home settings~\cite{soubutts2023:shifting}. Barbarin~\etal present that leveraging time-based objects such as pill boxes has benefits not only for better time management on medical activities but also for increasing family members' awareness of the patients' health activities in home care settings~\cite{barbarin2015:taking}.

\subsection{Technologies for Improving the Well-being of PwD and Caregivers}
\label{sec:oldtech}

Guan~\etal~\cite{guan2021:cue} suggest two types of caregiver-support technologies,\textit{mediating technology} and \textit{task-assistive technology}, focusing on PwD caregivers' relationship with care-receiver and caregiving activities. \textbf{Mediating technology} refers to tools and systems designed to foster and facilitate positive interaction between caregivers and PwD, serving as an interaction bridge. Some existing studies can be mapped as mediating technology. For example, Mu\~{n}oz~\etal and Unbehaun~\etal, respectively, designed video-game applications to promote positive social interactions between PwD and caregivers~\cite{munoz2021:app,unbehaun2020:appropriation,unbehaun2018:facilitating}. Houben~\etal presented smart home devices, \textit{Tumbler} and \textit{Vita}, that can play everyday sounds (\eg~sounds of the wind, a harbor, or a park) to evoke memories that can be used as the basis for conversation topics~\cite{houben2022:sound,houben2020:role}. Designed for therapeutic purposes, Zoomorphic robots (\eg~Sanne (cat-shaped)~\cite{marchetti2022:petrobot}, AIBO (dog-shaped)~\cite{tamura2004:entertainment}, Paro (seal-shaped)~\cite{marti2006:socially}), can help relieve psychological symptoms and induce positive emotions in PwD.

On the other hand, \textbf{task-assistive technology} aims to streamline the initiation and execution of caregiving tasks, making them more manageable for the caregiver. For instance, Wang~\etal proposed an assistive robot to support PwD's daily activities, such as making a cup of tea and handwashing, by prompting step-by-step guidance to complete tasks~\cite{wang2017:daily}. Bewernitz~\etal designed an assistive device to help with brushing teeth and upper body dressing by prompting messages~\cite{bewernitz2009:feasibility}. Many commercial technologies such as monitoring PwD through sensors and screens~\cite {sriram2019:informal} and reminding caregivers through multi-sensory modalities~\cite{cathrine2016:tracing} can be mapped here as well.

In contrast to the types of support previously discussed in~\autoref{rl:framework}, many of the technologies discussed in this~\autoref{sec:oldtech} directly intervene in caregiving situations to support caregivers' tasks. We therefore capture this type of support as \textbf{in-situ support}, which the prior framework~\cite{chen2013:integrality, schorch2016:designing} has not covered. In designing technologies for PwD caregivers, we need to holistically consider their different needs for support (social, information, coordination, and in-situ). Based on interview findings with 11 informal caregivers of PwD, therefore, our work aims at identifying design opportunities by situating such concerns within predictability and temporal contexts of verbal agitation from PwD.

\section{Methodology}

Our goal was to understand the impact of verbal agitation from PwD on informal caregivers and explore ways that technology may better support informal caregivers during (and around) these events. We conducted a semi-structured interview study with 11 informal caregivers. In particular, we explored the following research questions:

\begin{itemize}
    \item \textbf{RQ1}: In what exact situations do informal caregivers experience verbal agitation of PwD?
    \item \textbf{RQ2}: How do informal caregivers of PwD cope with the situation of verbal agitation?
    \item \textbf{RQ3}: What kind of technological support could be useful for the mental well-being of informal caregivers experiencing verbal agitation of PwD?
\end{itemize}

\subsection{Recruitment and Participants}
\label{subsec:participants}

We recruited participants (\ie informal caregivers of PwD) from South Korea and the USA, primarily to observe more diverse situations, rather than to focus on cultural differences. In doing so, we tried to avoid cultural (mis)attribution bias: that is, either overemphasizing or underemphasizing the impact of culture on the behaviors of participants depending on where they are from~\cite{jose2019:bias}. Despite the potential difficulties contacting sensitive target populations, we were fortunate to have connections to local dementia support communities in South Korea and the USA. Thus, we decided to recruit participants from these two countries.

For recruitment, we distributed flyers to local clinics, local dementia support groups, and dementia caregiver support communities (\eg libraries~\cite{dai2021:library}) both in Korea and USA. To be eligible for our interview study, individuals were required to meet the following conditions: i) be 18 or older, ii) be an informal caregiver of a person with dementia, and iii) have experienced verbal agitation, including screaming, swearing, threatening, blaming, repeating phrases, or making persistent noises such as moaning~\cite{nagaratnam2003:screaming,talerico2002:aggression,tueth1995:dementia,cariaga1991:vocalization}. We did not require cohabitation (see the cohabitation column in ~\autoref{tab:interviewees}). We screened participants' eligibility using an online survey before scheduling their sessions. Participants who completed a one-on-one online interview received a gift card with the amount of \$20 in their currency as compensation. Our Institutional Review Board (IRB) reviewed and approved this study.

We recruited 11 interviewees (2 males and 4 females from Korea; 3 males and 2 females from USA) who currently are or have been informal dementia caregivers, as summarized in \autoref{tab:interviewees}. All informal caregivers who participated in our interview study are familial caregivers who are either children or a spouse of PwD. We had two married couples: (P3, P4) and (P9, P10). Also, we had four participants (P5, P6, P9, P10) whose loved ones with dementia passed away within the past two years. None of the loved ones with dementia reside in any type of care center except P11. His mother had cohabited with his father until she moved to a dementia care residence a few months prior to the interview. The caregivers' ages ranged from 31 to 76 and the loved ones' ages ranged from 71 to 99. The periods for dementia ranged from 2 years to 15 years. Among the loved ones of the interviewees, six were diagnosed with Alzheimer's Disease (AD) (\aka senile dementia), one had vascular dementia, and the other four had mixed dementia (Alzheimer's and vascular).

\begin{table*}

\caption{Participant background. As informal caregivers, they all suffered from verbal agitation of their loved ones with dementia. In the \textit{Dementia Type} column, AD denotes Alzheimer's Disease. *People with dementia who did not cohabit with their caregivers lived in their own homes separately. **Periods of dementia.}
\label{tab:interviewees}

\centering
\resizebox{\textwidth}{!}{
\begin{tabular}{l l l l l l l l c l}
\toprule
      & & \multicolumn{3}{c}{\small \textit{Informal Caregiver}} & \multicolumn{3}{c}{\small \textit{People with Dementia}} \\ \cmidrule(lr){3-5}\cmidrule(lr){6-8}
    {\small \textit{Country}}
    & {\small\textit{\#}}
    & {\small \textit{Age}}
    & {\small \textit{Gender}}
    & {\small \textit{Job Status}}
    & {\small \textit{Relationship}}
    & {\small \textit{Age}}
    & {\small \textit{Dementia Type}}
    & {\small \textit{Cohabitation*}}
    & {\small \textit{Periods**}}\\
    \midrule
    \multirow{6}{*}{Korea}
          & 1 & 63 & Female & Full-time & Mother-in-law & 89 & AD & Yes & 4 years \\
          & 2 & 67 & Male & Not working & Father & 97 & AD & Yes & 9 years \\
          & 3 & 54 & Female & Not working & Mother-in-law & 94 & AD \& Vascular & Yes & 15 years \\
          & 4 & 68 & Male & Not working & Mother & 94 & AD \& Vascular & Yes & 15 years \\
          & 5 & 55 & Female & Full-time & Mother-in-law & 86 & AD & No & 8 years \\ 
          & 6 & 55 & Female & Freelancer & Father & 80 & Vascular & No & 6 years \\ 
    \hdashline
    \multirow{5}{*}{USA}
        & 7 & 57 & Female & Full-time & Mother & 77 & AD & No & 2 years \\ 
        & 8 & 76 & Male & Not working & Spouse & 77  & AD & Yes & 6 years \\ 
        & 9 & 73 & Male & Freelancer & Mother-in-law & 99 & AD \& Vascular & No & 8 years \\ 
        & 10 & 67 & Female & Full-time & Mother & 99 & AD \& Vascular & No & 8 years \\ 
        & 11 & 31 & Male & Full-time & Mother & 71 & AD & No & 6 years \\ 

\bottomrule
\end{tabular}
}
\end{table*}

\subsection{Interview Protocols}
\label{subsec:interview}
We conducted one-on-one semi-structured interviews through Zoom with informal caregivers of PwD. At the beginning of the interview, we disclosed that the interviewer (\ie the first author) and another member of our research team (\ie the second author) are also informal caregivers who have suffered from verbal agitation of PwD. This is to prevent interviewees from feeling embarrassed and to encourage their willingness to share their sensitive experiences~\cite{murphy1972similarity, elmir2011interview}. Next, we explained that the purpose of the study is to understand the challenges and technological needs to support informal dementia caregivers experiencing verbal agitation of PwD. We had interviews in the participants' primary language (English or Korean). We audio recorded with the interviewee's consent and allowed participants to choose between audio or video interviews. Each interview took about an hour. We further discuss how we addressed several ethical considerations of our study in \autoref{sec:ethical}.

We asked questions that cover i) caregivers' experiences with verbal agitation, ii) how the interviewee currently copes with mental distress posed by those burdens and situations, and iii) what technological support would be helpful to facilitate their mental well-being. We also asked a few questions about their basic information, including the caregivers' gender, job status, relationship with the loved one, the age of the loved one, types of dementia, and periods of care as presented in ~\autoref{tab:interviewees}. The language for the seed questions for the semi-structured interview was carefully reviewed and revised by two scientists in our research team: [A] a professor with 25 years of clinical experience with older adults with neurodegenerative disorders, and [B] a professor in preventive medicine. To increase the clarity and validity of the responses, we encouraged participants to come up with tangible examples (\eg ``Last Friday, she told us `get the f**k outta here, b**ch!', when we tried to change her diaper.'') rather than abstract descriptions (\eg ``She often becomes nasty to me and my wife, when we tried to help her go to bed in the evening.'') We asked participants to try to look back at the recent two weeks for helping them bring up concrete examples.

\subsection{Data Analysis}
\label{sec:analysis}
The first author transcribed all the audio recordings and anonymized identifiable data, such as people's names, in the transcripts. For interviews with Korean participants, the first author, who is bilingual in both Korean and English, translated the transcripts into English for data analysis. We merged transcripts of Koreans and Americans to analyze them together. That is because we intended to have inclusive and diverse findings from diverse settings, rather than comparative findings between the two groups, as noted in~\autoref{subsec:participants}. Upon the completion of the data preparation, three of the authors conducted thematic analysis~\cite{braun2006:thematic} on the merged transcripts of all interview recordings. We first open-coded each transcript independently to distill the initial codes with research questions in mind. Then three authors had regular meetings to discuss and reshape themes to reach an agreement on potential themes. Through several rounds of discussion, we carefully reviewed and refined themes to deliver a coherent story by merging some themes into higher-level categories, dividing other themes into separate categories, and discarding themes not relevant to our core research questions. In the end, we determined the final themes for each research question.

\subsection{Ethical Considerations}
\label{sec:ethical}

We had to consider a number of ethical issues throughout our study. That is because the interview questions investigated participants' personal experiences with their loved ones' (\ie~PwD) verbal agitation that would often entail mental distress. Two professors (\ie~[A] and [B] mentioned in~\autoref{subsec:interview}) in our research team taught and trained interviewer (\ie~the first author) in the required knowledge and interview skills appropriate for our particular study population (\ie~dementia caregivers) based on the following considerations. First, we used person-first language~\cite{sharif2022:peoplefirst} -- \textit{people (person) with dementia} instead of \textit{dementia patients} -- throughout the interviews, because we felt that referring to a caregiver's loved one as a \textit{patient} may make them feel uncomfortable or trigger mental distress. Secondly, to avoid any confusion, we clearly informed participants that the purpose of our study is to investigate how future technologies may better support informal dementia caregivers' mental health, but not to provide immediate clinical suggestions during the interview. Thirdly, we monitored their emotional state during the interview by i) checking both verbal and nonverbal cues from the interviewees, and ii) directly asking them if they were fine or willing to continue. None of our participants indicated emotional discomfort during the interview. Lastly, we continuously reminded participants that they do not have to answer any questions and may stop the interview at any time if they feel uncomfortable.

\section{findings}

We find that informal caregivers experience verbal agitation of PwD in diverse contexts and their technological needs vary by the timing of such behaviors. First, we characterize contexts of verbal agitation of PwD in terms of \textit{predictability} to informal caregivers and \textit{target direction} (\eg~informal caregivers or third party). Then, we present our participants' coping strategies and their proposed technological support before, during, and after verbal agitation. \textbf{[Trigger Warning]} The findings and quotes below may contain cursing, aggressive language, and suicidal ideation. We masked quotes that included explicit curse words.

\subsection{Overview of Impacts of Verbal Agitation on Informal Caregivers' Mental Health}

Echoing previous work~\cite{draper2000:casecontrol}, participants described substantial negative impacts on their mental health due to verbal agitation of PwD (\eg~cursing at informal caregivers, repeating the same questions over a three-hour period). Some participants mentioned that they went through serious suicidal thoughts (P3) or even prayed for their loved one with dementia to die (P10). In addition, participants reported impacts on their physical and social health. They experienced manifestations of stress such as significant weight loss, insomnia, and high thyroid hormone levels. False rumors (\eg~``He stole my money. They always gave me shitty foods.''--P2) that PwD spread around relatives and neighbors tarnished participants' social relationships and reputations. Participants noted that these impacts on physical and social health exacerbated their mental distress, as well.

\subsection{Predictability and Target Direction of Verbal Agitation of PwD (RQ1)}
\label{find:rq1}

Participants described \textit{unpredictable} verbal agitation for which they could not identify triggering situations, as well as \textit{predictable} verbal agitation that happened repeatedly under similar contexts. Participants also experienced various target directions of verbal agitation, including toward family members, informal caregivers themselves, or a third party.

\subsubsection{Unpredictable and predictable contexts}
\label{find:predict}

Our participants experienced verbal agitation that occurred during one-off events that were \textit{unpredictable}. In these contexts, participants commonly acknowledged that they were mentally vulnerable and emotionally insecure because verbal agitation happened out of the blue. For example, P1's mother-in-law abruptly yelled at P1, while P1 was cooking in the kitchen and her in-law was sitting on the couch in the living room. P1 thought her in-law was taking a rest. Since they were not even in any kind of interaction or conversation, P1 was startled when her in-law manifested such verbal agitation.

\begin{quote}
P1: She [mother-in-law] cursed at my husband like ``You are a f**king maggot!!'' That's way too much all of a sudden. It was really shocking.
\end{quote}

On the other hand, participants were often \textit{able to predict} verbal agitation that repeated in certain routines, like personal care (\eg~taking a shower), traveling by car, and visiting public spaces (\eg~restaurants). We originally presumed that participants would be less stressed out in predictable contexts because predictability would reduce the shock by allowing them to be mentally and physically prepared. However, they said that, rather than being useful, such predictability was mentally exhausting. Being able to anticipate upcoming verbal agitation did not help them to prevent it from happening, but it made them imagine an impending unpleasant experience, increasing their mental burden. For example, P1 reported that changing PwD's diapers always led to verbal agitation, and P3 felt mentally exhausted whenever she needed to leave her loved one with dementia for personal errands. Awareness of such conditions or cases triggering verbal agitation led them to anticipate a negative situation and hence made them stressed.

\begin{quote}
P1: She [mother-in-law] cursed at me when I tried to change her diaper. She screamed ``Look at this b**ch! This b**ch is taking off my underwear!!'' I had to go through this \textbf{whenever changing her diaper}.
\end{quote}
\begin{quote}
P3: Maybe she [mother-in-law] thought that I abandoned her \textbf{whenever I leave her for any reason}. She gets angry, saying ``Am I a housekeeping dog? am I a shepherd!?'' She never allows me to leave home.
\end{quote}

Given these differences in participants' responses to verbal agitation events, therefore, support mechanisms may need to aim at relieving the shock of an unexpected situation and overcoming anticipatory anxiety for an expected situation, for example. We discuss them further in the design opportunities (\autoref{sec:design}).

\subsubsection{Target direction of verbal agitation}
\label{find:direction}

Participants perceived different types of mental distress depending on the target audience group of verbal agitation from PwD. Some verbal agitation was directed at strangers or a third party (\eg P5) and others were targeted at informal caregivers. For example, P5 described her experience in a restaurant with her mom.

\begin{quote}
P5: Sigh... it was truly embarrassing. She [mom] shouted, ``How dare such green as grass b**ches have dinner in this restaurant!'' I had to go up to other tables to apologize, saying my mom was very sick.
\end{quote}

While verbal agitation directed at strangers made them flustered and perplexed, those targeted to informal caregivers caused significant mental distress to them. For example, P3 described that she had suicidal thoughts when her mother-in-law cursed at her when she returned home from the grocery store. Given that the target direction of verbal agitation could have qualitatively different impacts on informal caregivers, designers and researchers need to investigate interventions that take the audience of verbal agitation into account.

\subsection{Current Coping Skills before/during/after Verbal Agitation (RQ2)}
\label{find:rq2}

We found that participants apply different coping methods depending on timing relative to an event: before, during, or after verbal agitation. 

\subsubsection{Impending verbal agitation (pre-event)}
\label{find:rq2pre}

Participants reported their coping strategies when they perceived an impending verbal agitation (but before the verbal agitation actually began). First, some participants tried to physically get away from PwD. For example, P3, P4, and P11 said they just walked outside of the house to avoid conflict:

\begin{quote}
P3: I just get out of my home. I walk outside. I try my best to quickly get away from \textbf{the impending moment} because I know that I should not begin arguing [with my mother-in-law].
\end{quote}

Secondly, some other participants often distracted PwD by giving them different topics or objects to focus on. For instance, P1 said that she turned on the TV to distract her loved one with dementia, and P8 changed the topic to distract his wife with dementia:

\begin{quote}
P8: I try to work around [the impending moment], you know. I change the [conversation] subject to distract her in some way. Maybe leave it for the moment if I can, and then come back to it.
\end{quote}

Lastly, some participants have been deceiving PwD to make them feel comfortable. For example, P7 had to deceive her mom with dementia. P7 and her sister knew that their mom would have manifested verbal agitation (\eg yelling, persistently complaining) towards P7 and her sister if she had found that they had called the nurse to help their mom with toileting. Thus, they told their mom that they did not call the nurse, even though they had:

\begin{quote}
P7: It's sneakier but less upsetting [for mom]. When my sister and I get the nurse, then my mom just resigns. She [mom] just accepts the situation.
\end{quote}

These strategies for avoiding or preventing verbal agitation apply largely to cases of predictable agitation. However, informal caregivers would not be able to avoid \textbf{un}predictable verbal agitation (\autoref{find:rq1}) beforehand, and they may not be able to physically distance themselves from PwD at the very moment of a verbal agitation event. Therefore, they adopt different methods for coping with verbal agitation during events, discussed next.

\subsubsection{During verbal agitation (within-event)}

Participants described several strategies they have applied during verbal agitation. One common technique was to respond to every verbal agitation (conversation). For example, P6 said that she and her mom took turns to keep responding to her dad repeating the same question for over three hours, and that it was both mentally and physically exhausting. However, responding to every verbal agitation may not be an ideal way to handle it, as informal caregivers are not always able to find someone who can take turns with them.

\begin{quote}
    P6: My mom and I anyway tried to respond to his [dad's] endless repeating questions over and over again no matter how many times he asked. But it really drove us crazy. We lost our voice afterward.  
\end{quote}

Some other participants tried to go along with verbal agitation, like complaints, to avoid a potential struggle or dispute. For example, P11 described that his mom cursed and complained at a public restaurant, saying she wanted to have ice cream, not fries. However, his mom had indeed ordered fries following the cashier's recommendation, but his mom argued that the cashier forced her to order fries. In this context, P11 responded as follows:

\begin{quote}
P11: The best response would be ``Yeah damn you're right! I'll put these aside. We're going to have the ice cream!'' I've learned how to throw away all expectations which allows for less power struggles. Just allow whatever was planned to not go as planned and be okay with it.
\end{quote}

Many participants acknowledged that they often lost the ability to be nice and agreeable due to mental and physical exhaustion. For instance, many participants said they failed to apply this `going with the flow' strategy because they were often unable to control their emotions, even though they were aware that this strategy was better than confrontation to not aggravate the situation (\eg~P7, P8, P10). This reality motivates exploring opportunities for technology that helps informal caregivers to manage their own temper, so they can apply more effective strategies to eventually calm down PwDs' verbal agitation.

\subsubsection{After verbal agitation (post-event)}
\label{find:post-event}

The time periods after verbal agitation can be identified as \textit{proximal} and \textit{distal}. The former means a relatively short time period immediately after verbal agitation while the latter refers to a longer time period between the latest verbal agitation and the forthcoming event. \textit{In proximal contexts}, participants focused on recovering their emotional damage and mental distress caused by verbal agitation. Some participants utilized media content such as music and movies as a remedy. They also looked for close friends or family members to complain to about verbal agitation and receive comfort from. For example, P1 and P3 turned on music that can help them vent their fury or calm down for meditation. P1 added that she often maximized the volume of music to block the sound of verbal agitation from PwD:

\begin{quote}
P1: [Immediately after verbal agitation] I often turn on the music super loudly to represent my emotion. I listen to some Kpop, Adele, and Whitney Houston songs. I believe that [listening to music] is by far the best way to manage my emotion effectively.
\end{quote}

However, they often found it difficult to share their complaints with friends and other family members because they might not share the same understanding of their contexts regarding dementia. Also, they were often frustrated when some friends or family members failed to appreciate or acknowledge informal caregivers' competence and the effort they made for care activities (P3).

On the other hand, participants' \textit{distal coping skills}, between verbal agitation events, often included preparations for future verbal agitation. Some of them went to a local caregiver support group to learn better care practices. They studied proper coping skills on the internet. For example, P8 took an online course program to learn more about dementia:

\begin{quote}
P8: I learned by myself. and I took an online course. It's a 6-week program to complete. Also, it's been over 2 years now going to meet people at the caregiver support group.
\end{quote}

Lastly, to improve their mental health, some participants (\eg~P2, P11) commonly pursued activities that could give them a greater sense of autonomy. For example, they enjoyed working out (\eg~walking, weight lifting, cycling, tennis) and socializing (\eg~chitchat, drinking alcohol). Given that informal caregivers' freedom is largely restricted in the dementia care context~\cite{sanders2008:grief}, it is necessary to consider various ways to restore their freedom in designing support systems for their mental health.

\subsection{Participants' Needs Based on their Suggested Ideas for Mental Health Support (RQ3)}
\label{find:rq3}

We describe participants' needs based on their proposed ideas for pre-/in-/post- verbal agitation of PwD. We elicited types of mental health support that our participants looked for by asking them to imagine potential technologies (RQ3). To prompt this, we asked questions like \textit{``Suppose there are several movie clips of verbal agitation that you experienced with your PwD, and you are the director of those films. You can do anything with the film. How do you want to change or edit those films to protect or support the caregiver's mental health in the films?, if you have a magic wand, how do you want to help?''} These questions allowed participants to imagine speculative systems to support their mental health. With participants' suggested ideas, we could achieve a better understanding of problems that they care more about and how systems may integrate into their lives. In this section, we describe both potential technologies imagined by participants as well as the underlying needs we believe those technologies are intended to address.

\subsubsection{Having space to prepare for the impending moment (pre-event)}
\label{tech:pre1}

Our participants mentioned that they needed support to help them mentally and physically prepare for an impending verbal agitation. They suggested two possible ways to achieve this. First, intelligent systems may detect and \textbf{alert impending verbal agitation} by sensing bio-signals or voice tones of PwD. P11 said that informal caregivers would not always be able to monitor PwD, even in home settings, because they could be focusing on other tasks such as cooking and cleaning. Then, he specified that video (image, animation) or audio (music) cues could reflect his PwD's vitals and signal impending verbal agitation events. Secondly, systems could buy informal caregivers some time to prepare by \textbf{presenting media to distract PwD} when an oncoming verbal agitation from PwD is anticipated. P11 suggested utilizing a hologram to distract PwD's attention to prevent impending verbal agitation:
\begin{quote}
P11: Maybe Alexa can give some distraction. If a hologram [of my mom's favorite characters or animals] came up in front of her [mom] She'd be very nice like ``Hi there, what are you doing here boy?''.    
\end{quote}

Alert and distraction systems can help informal caregivers by giving them time to prepare for wrapping up their personal tasks and getting ready before shifting their focus to impending verbal agitation. Therefore, our participants want to have space to prepare for a smooth transition between personal life and care activities, as it helps them prevent burnout.

\subsubsection{Securing self-control when it happens (within-event)}
\label{tech:mid1}

Our participants showed that they want technology that can facilitate securing their own self-control during verbal agitation of PwD. To achieve this goal, they wanted to be able to avoid hearing aggressive verbal expressions and tones. Specifically, they suggested a voice assistant that \textbf{filters out the provocative aggression} of verbal agitation in real-time may help informal caregivers secure self-control by alleviating the impact on them. For example, a system could paraphrase the aggressive content to make it sound pleasant. Or, it could partially or entirely mute the aggressive content (\eg~swearing) to avoid the negative impact on mental health. However, P6 noted that she wants to listen to what her dad said to her, but without offensive words to understand any discomfort that should be addressed in the moment. Therefore, she wanted technology that refines expressions or word choices to sound less offensive, or even sound ``beautiful'', while retaining the ability to understand what her dad is saying without her getting upset. Similarly, P8 wanted his hearing aids to partially or even completely mute some offensive words from his wife. He explained that he could not help getting upset, for instance, when his wife insisted on not getting out of the car in the parking lot. Thus, he often did not want to listen to words that made him get annoyed. 

\begin{quote}
P6: I wanted a system to filter out or rephrase her words. Maybe something that can refine her expressions more nicely and beautifully.
\end{quote}

Secondly, to have themselves secure self-control, participants wanted tools that can support \textbf{improving self-awareness of their own emotions} when they are experiencing verbal agitation of PwD, such as a system that helps monitor their biosignals such as heartbeat, blood pressure, and body temperature to be aware of their emotional status. P8 explained that he always wanted to be nicer to his wife even during verbal agitation, because he knows that his wife cannot control herself and she does not mean to manifest verbal agitation. However, he naturally felt angry in the moment of verbal agitation from his wife. This made him helpless. In this regard, he wished that an electric shock could be administered via his hearing aid to remind him to have better self-awareness of his anger. While we do not condone designs using electric shocks, the severity of this participant's design suggestion demonstrates their intense need for securing self-control during a verbal agitation event:

\begin{quote}
P8: Maybe something where I'm wired to an electric shock when I start getting annoyed so that I won't be mad. 
\end{quote}

Our participants suggested systems to reduce or avoid the negative impact of verbal agitation. Also, they proposed systems to have better self-emotion awareness during verbal agitation. In general, this suggests a need for technological support to enable them to stay sharp and keep their self-control during verbal agitation of PwD.

\subsubsection{Receiving emotional support (post-event)---proximal}
\label{tech:end1}

We define the proximal period as the moment immediately after verbal agitation of PwD (\autoref{find:post-event}). For this period, participants wanted technology to comfort them to ameliorate their negative emotional state and relieve distress. To achieve this goal, they suggested \textbf{conversational technology} that can verbalize empathic words and acknowledge their challenging situations. Participants noted that they often did not expect solutions, but a rather proper acknowledgment of the situation and the emotion, especially immediately after verbal agitation. For example, 

\begin{quote}
    P3: When I would say ``Do I really have to take care of her even in this crazy situation? What do you think?'', ``Isn't she supposed to be in the nursing residence for dementia care?'', then if the robot responded like ``I completely agree.'', ``Yes, she's supposed to be in the residential care center, not here in your place.'', these would be very stress-relieving responses for me. 
\end{quote}

Our participants suggested another system that can \textbf{play media content} to improve their mental health. They mentioned playing music for venting out their emotions (\eg~pop music--P3), and for calming down and relaxing themselves (\eg~new age piano songs--P1). Also, they considered other types of media including images, videos, etc. For example,

\begin{quote}
P11: In home, some relaxing music comes on, a familiar photo pops up on the wall. You know, I can see that a little bit some you know that would be kind of helpful and nice.
\end{quote}

The participants desire technology that they can use to confide in and receive emotional support through conversation and media. That is to say, they want to get appropriate comfort to recover from negative emotions and mental distress in the proximal moment after verbal agitation.

\subsubsection{Preparing for future verbal agitation (post-event)---distal}
\label{tech:end2}

For this distal period, participants wanted technology focusing on preparation for future verbal agitation of PwD. Therefore, they wanted technology to \textbf{provide learning materials}, including better dementia care practice, more potential symptoms of dementia, and coping skills for mental self-care. Our participants wanted to learn better care practices because they had not received any professional training as a dementia caregiver. Also, they wanted to learn more about potential symptoms of dementia through media content.

However, in addition to informal caregivers' general lack of access to educational resources about dementia~\cite{dixon2022:barriers}, our participants often found those materials were difficult to understand and imagine dementia symptoms. For example, a terse description that verbal agitation includes screaming, cursing, etc. is not sufficient for them to make use of in practice. Instead of such simple descriptions, they wanted a video clip of real examples of dementia symptoms. Therefore, participants wanted a system to provide training materials that they could use for better care practices. 

\begin{quote}
P5: On the Internet, you know, it says like -- \textit{verbal agitation such as cursing, repeating words, ... blah blah is a symptom of dementia}. But those descriptions are way too abstract. I want a system that can provide a list of examples of symptoms. Not just descriptions but with images, videos, and voices maybe. Knowing that `this is not just my PwD's symptom [but it is a common symptom]' is very relieving.
\end{quote}

Another idea that several participants wanted to have is \textbf{a handy simulation training} to prepare for future verbal agitation. They wanted a system to help them apply better practices in their daily care context. Systems simulating verbal agitation would allow informal caregivers to practice coping skills that they have learned from somewhere. For example,

\begin{quote}
P8: One thing that would be kind of handy is role-playing simulations. It says like ``Would you mind saying this instead or doing this.'' It basically is training you how to circumvent the situation, how to sidestep which is really what you need to do.
\end{quote}

As we noted, many informal caregivers inevitably face unpredictable verbal agitation of PwD (\autoref{find:predict}). In irregular and abrupt contexts, informal caregivers may not have enough time to mentally get familiar with the situation. Such simulation-based training would enable informal caregivers to be more prepared for future verbal agitation.

\subsubsection{Form factors of technology}
\label{find:form}

Several participants brainstormed the form and voice features of potential technologies. P7 first raised privacy concerns about the visibility and size of a system. She said that the absence of a physical form that demonstrates the existence of an agent (\eg~Siri in iPhone, Alexa in Echo) would be concerning regarding information privacy.
\begin{quote}
P7: If I wear a watch talking to me, I wonder: is it recording me, who does send this message, how is the message determined, is it for me, and so on. It will feel invasive.
\end{quote}
She preferred a robot that is sufficiently large to be perceived as an agent. A robot could provide physical support for her mom with dementia as well as mental support for her. However, she was concerned about her physical privacy if the robot is too large (``six feet tall''), because it takes up too much of her space. Her concerns align with privacy considerations in Human-Robot Interaction design~\cite{lutz2019:privacy}. Both informational and physical privacy, thus, should matter for designing systems to improve the mental health of informal caregivers.

Participants also suggested the system have the voice of a person with whom they would love to talk. For example, some participants suggested their granddaughters' voices (\eg P8) or the voice of an actor (\eg P6, P10).
\begin{quote}
P10: In Star Wars, Ewan McGregor? Oh my gosh his voice, I could listen to that forever.
\end{quote}
Given that listening to a familiar voice help people to calm down their unhappy emotional state~\cite{grossbach2011:ventialtion}, an agent that speaks in the voice of an informal caregiver's family member would help them to relieve their mental stress.

\section{design opportunities}
\label{sec:design}

\begin{figure*}
    \centering
    \includegraphics[width=\textwidth]{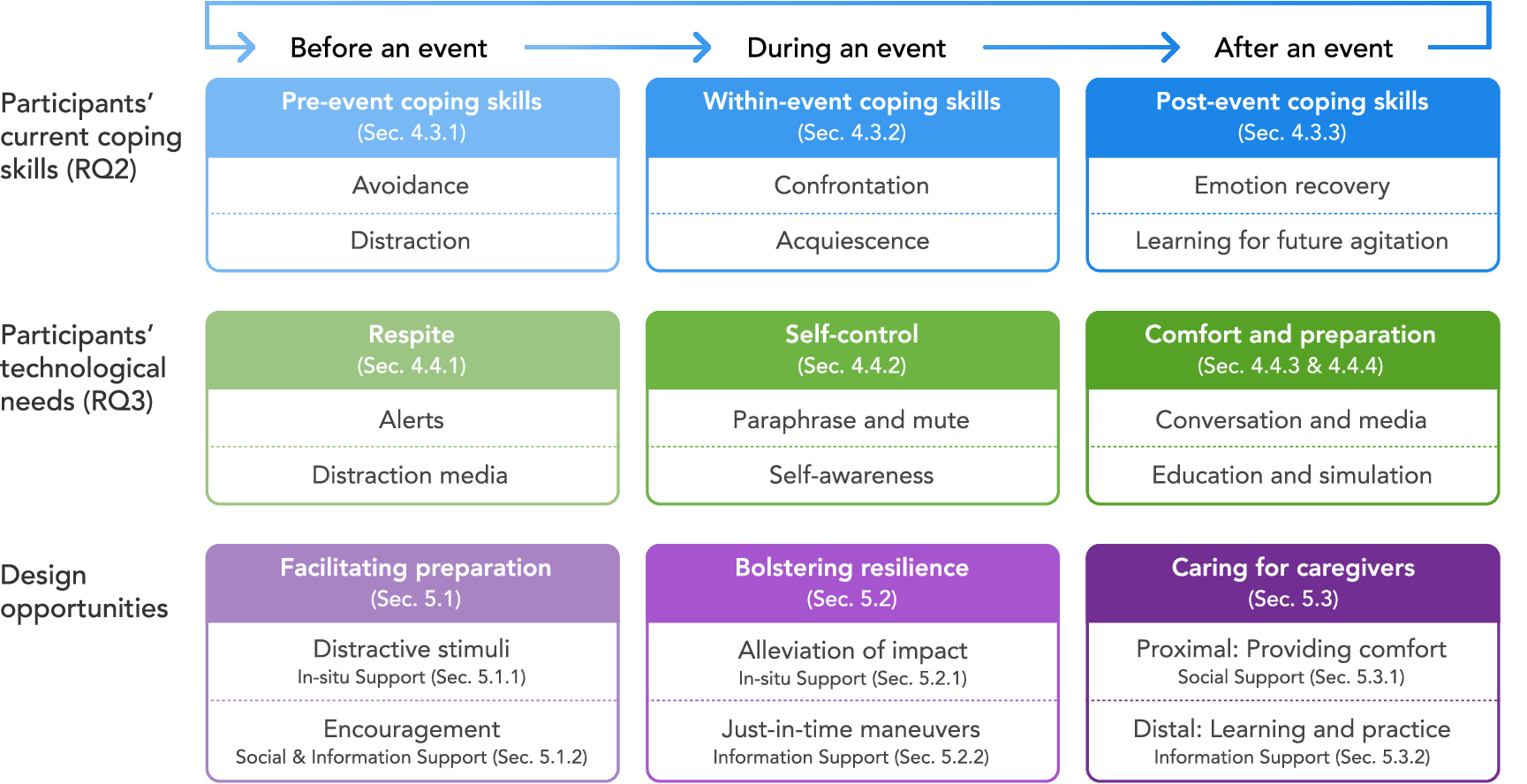}
    \caption{We synthesize our findings on current coping strategies (RQ2) and technology suggestions (RQ3) of our participants (informal caregivers) to derive design opportunities for before/during/after verbal agitation.}
    \label{fig:overview}
\end{figure*}

Predictability of verbal agitation impacts when and how informal caregivers feel burdened and exhausted (\autoref{find:rq1}). In predictable contexts, informal caregivers might be exhausted by recurring verbal agitation in the course of essential and unavoidable duties, such as toileting, eating, and dressing. On the other hand, unpredictable verbal agitation can make informal caregivers mentally vulnerable, struggling to form an appropriate response at the moment. However, predictability is not fixed, because dementia symptoms change over time. For example, verbal agitation becomes more definite in later stages of dementia~\cite{parmelee1993:pain,hurley1992:discomfort}, so it takes time for caregivers to be able to predict verbal agitation until they get used to those provoking conditions. In addition, we found that our participants needed different technological support before, during, or after verbal agitation. For instance, they desired tools to support their self-control during verbal agitation, while they needed comfort technology for their emotions after the event (\autoref{fig:overview}).

Below, we discuss design opportunities in terms of predictability and timing of verbal agitation. Instead of reproducing the participants' responses in~\autoref{find:rq3}, we aim to reimagine and repurpose our findings into design opportunities, using a broader lens of research and practical applications. We further outline how design opportunities we characterize can sync with the existing caregiver support framework (\textit{social support},~\textit{information support}, and \textit{coordination support}~\cite{chen2013:integrality, schorch2016:designing}) and our proposed fourth category of support, \textit{in-situ support} (\autoref{sec:oldtech}). First, we explore how to facilitate informal caregivers' preparation \textit{before predictable} verbal agitation. Next, we present design opportunities for bolstering self-control \textit{during unpredictable} verbal agitation. Lastly, we discuss providing comfort \textit{after verbal agitation (proximal)} and supporting learning and practice \textit{after verbal agitation (distal)}. We tabulated several examples of potential applications, potential modalities, and artifacts to build those applications in \autoref{tab:design}. We do not propose design ideas for \textit{coordination support}. This type of support mainly relates to the management and planning aspects of an informal caregiver's routine~\cite{chen2013:integrality, schorch2016:designing}. Although previous research offers various suggestions for \textit{coordination support} like shared calendars~\cite{eschler2015:shared}, our interview data do not suggest that such types of support are what informal caregivers look for in the given context.

\subsection{Facilitating Informal Caregivers' Preparation before Predictable Verbal Agitation}

We identify two challenges in supporting informal caregivers' mental health before predictable verbal agitation. First, informal caregivers struggle with having space to prepare before engaging in the situation of predictable verbal agitation. We suggest technology that \textbf{distracts} PwD to help informal caregivers have space to prepare themselves before impending verbal agitation. Second, anticipatory anxiety due to the awareness of an upcoming verbal agitation makes informal caregivers feel helpless, deterring them from initiating certain care activities. To address this issue, we propose technology that \textbf{encourages} these caregivers to start their caregiving tasks.

\subsubsection{Distractive stimuli towards people with dementia (in-situ support)}
\label{dis:distraction}

We suggest technology offering distractions with different modalities to PwD before impending verbal agitation, lowering the chances of verbal agitation or even delaying it from happening. Distractive stimuli can be made through music~\cite{guan2021:cue}, physical objects~\cite{houben2022:sound,marti2006:socially,houben2020:role}, interactive activities~\cite{munoz2021:app,unbehaun2020:appropriation,unbehaun2018:facilitating}, art~\cite{cornejo2016:vulnerability}, photographs~\cite{piper2013:audio}, or conversations~\cite{boyd2015:beyond,houben2020:role}, all tailored to PwD's interest. These technologies could afford informal caregivers space to prepare themselves prior to impending verbal agitation.

\ipstart{Potential usage scenario} Andrew is the primary caregiver for his father, diagnosed with dementia. He knows that his father responds with intense denial, shouting, and verbal aggression whenever Andrew attempts to change his diaper. To address this challenge, Andrew purchased Paro, a seal-shaped intelligent robot~\cite{marti2006:socially}, that can play everyday sounds (\eg sound of his old dog) that can remind his father of past memories~\cite{houben2022:sound}. It can also play music and initiate conversations. Now, when changing a diaper, Andrew sets up Paro to play his father's favorite tunes that he used to sing a lot (\textbf{in-situ support}). While his father is interacting with Paro, Andrew is mentally prepared and able to commence with the diaper change.

\ipstart{Design considerations} Designing distractive stimuli needs to consider modality and overuse. Informal caregivers may need to operate a distraction device with different modalities at their own convenience. When their hands are occupied by care tasks, such as changing diapers,  voice commands may be one alternative modality to operate systems~\cite{chang2019:howto}. Future work can study appropriate modalities for distraction technologies. On the other hand, such distractive technology may pose ethical concerns. For example, some caregivers might overuse a distraction device, leaving the PwD constantly distracted by it and thus ignoring their care responsibilities. A joint engagement system (\eg \cite{unbehaun2020:appropriation,unbehaun2018:facilitating}) that requires participation from both caregivers and PwD could be a way to avoid such potential misuse.

\subsubsection{Encouragement for informal caregivers (social support \& information support)}

Encouraging informal caregivers to initiate and complete care tasks before predictable verbal agitation could boost their confidence and willingness. Encouragement technology could have various forms: smart devices (\eg~earbuds, smart phones) for auditory reminders~\cite{hoggan2009:audio,mcgee2011:user} or text prompts~\cite{kocielnik2017:send} and wearable devices (\eg~smart watch) for haptic signals~\cite{vaucelle2009:design}. As well as such motivational encouragement, informative support (\eg~providing guidance on care practices~\cite{guan2021:cue}) could encourage informal caregivers, depending on their individual preferences.

\ipstart{Potential usage scenario} Ben is taking care of his mother with Alzheimer's disease. Whenever he helps her brush her teeth, she shouts at him. This predictable reaction makes Ben overwhelmed and hence hesitant even before taking her to the bathroom. To cope with this, Ben utilizes his earbuds and smart watch. The watch, using biosignals like heart rates, activity tracking, and movement detection, interprets Ben's situation, and sends a haptic alert. Then, Ben checks his watch screen displaying motivational messages---``\textit{Ben, you got this. Your mom still loves you. You know, she does not mean it.}''--- from the online forums he frequently visits for peer support (\textbf{social support}). Shortly later, his earbuds play remedies to reduce conflict (\textbf{information support}). This encouragement technology helps Ben to manage care tasks with enhanced confidence and reduced stress.

\ipstart{Design considerations} Individual caregivers may have different preferences in receiving encouragement. For example, direct-minded people may prefer to be reminded of what is going to happen, while others may be better encouraged with empathetic messages. Thus, designing technologies for encouraging informal caregivers should be flexible with such individual preferences.

\begin{table}

\caption{A potential example of application, modalities, and artifacts for i) before, ii) during, and iii) after verbal agitation of PwD for improving informal caregivers' mental well-being.}
\label{tab:design}

\centering
\resizebox{\textwidth}{!}{
\begin{tabular}{l l l l l l}
\toprule

    & & \multicolumn{3}{c}{\small \textit{Design Opportunities} (\autoref{sec:design})} \\
    \cmidrule(lr){3-5}
    
    & {\small \textit{Support Type}} & {\small \textit{Application Example}} & {\small \textit{Artifacts}} & {\small \textit{Modalities}}\\
    \midrule
    \multirow{4}{*}{Before} & \multirow{2}{*}{In-situ} & \multirow{2}{*}{Distraction tools} & {Zoomorphic robots (\eg~Paro~\cite{marti2006:socially})} & \multirow{2}{*}{Audio, voice, video, image}\\
     &  &  & {and daily artifacts (\eg~Vita~\cite{houben2020:role})} & \\
     & \multirow{2}{*}{Information} & \multirow{2}{*}{Encouragements} & {Reminders through earbuds and} & \multirow{2}{*}{Audio, text, haptic}\\
     & & &  {alerts from smart watches} &\\
     
    \hdashline
    \multirow{4}{*}{During} & \multirow{2}{*}{In-situ} & \multirow{2}{*}{Refining provocative words} & {Filtering out swear words} & \multirow{2}{*}{Audio}\\
     & & & {and converting them to music} & \\
     & \multirow{2}{*}{Information} & \multirow{2}{*}{Reminding timely maneuvers} & {Visual prompts through AR device} & \multirow{2}{*}{Visual, text, audio}\\
     & & & {and text notifications for next steps} &\\

    \hdashline
    \multirow{4}{*}{After} & \multirow{2}{*}{Social} & \multirow{2}{*}{Comforting caregivers} & \multirow{2}{*}{Conversation-based stress relief} 
    & \multirow{2}{*}{Voice, audio}\\
    & & & \\
    & \multirow{2}{*}{Information} & \multirow{2}{*}{Reflecting care practices} & {Audio/video demonstration and} & \multirow{2}{*}{Video, AR, VR} \\
    & & & {simulation-based learning} & \\
    
\bottomrule
\end{tabular}
}
\end{table}

\subsection{Bolstering Informal Caregivers' Resilience during Unpredictable Verbal Agitation}
\label{dis:within_event}

Informal caregivers experience verbal agitation in unpredictable contexts, such as public spaces and in-home settings (\autoref{find:predict}). We identify two challenges for making informal caregivers retain their self-control in this context. First, informal caregivers lose their temper due to unpredictable verbal agitation. For this issue, we suggest technology that can \textbf{alleviate} the impact of verbal agitation during the event. Second, informal caregivers may feel overwhelmed and hence unsure of what to do when facing unexpected situations. To address this, we propose technology that \textbf{prompts} informal caregivers to stay focused and take timely actions. These technologies could help them build resilience and enable them to react properly in unpredictable situations. Informal caregivers experience verbal agitation in unpredictable contexts, such as public spaces and in-home settings (\autoref{find:predict}). We identify two challenges for making informal caregivers retain their self-control in this context. First, informal caregivers lose their temper due to unpredictable verbal agitation. For this issue, we suggest technology that can \textbf{alleviate} the impact of verbal agitation during the event. Second, informal caregivers may feel overwhelmed and hence unsure of what to do when facing unexpected situations. To address this, we propose technology that \textbf{prompts} informal caregivers to stay focused and take timely actions. These technologies could help them build resilience and enable them to react properly in unpredictable situations.

\subsubsection{Alleviating the Impact of Verbal Agitation on Listeners (in-situ support)}
\label{dis:para}
We motivate automated reshaping of verbal agitation to alleviate its impact on listeners and hence help them maintain their self-control during unpredictable verbal agitation. There can be different ways to reshape verbal agitation: completely eliminating curse words~\cite{bertero2016:real,md2023:hate} and converting offensive language to lovely words like song lyrics~\cite{kim2019:lyrics}. Such technology can be embedded in various devices, like smart earbuds or home speaker systems, making it accessible and effective in various care scenarios.

\ipstart{Potential usage scenario}
Charles has been taking care of his father with Alzheimer's disease for a decade. One day, they are on their way to the grocery store. Suddenly, his father started to curse at Charles---``\textit{You are f**king a**hole! You don't kidnap me!}''---while he is driving, which can be risky if Charles loses his temper. However, Charles is wearing earbuds with a swear-canceling feature embedded. The swear-canceling feature eliminates any words containing profanity (\textbf{in-situ support}). Hence, what Charles hears is simply, ``\textit{You don't kidnap me!}'' This feature allows Charles to drive safely without the distraction from harsh language.

\ipstart{Design considerations} 
PwD's verbal agitation often expresses their feeling of discomfort~\cite{kovach1999:assessment}. Reshaping technology could potentially make informal caregivers fail to sense their PwD's discomfort. Instead, designers could design technology that allows users to adjust the degree of refinement applied to aggressive language. For example, in situations where complete elimination of aggressive language is necessary, like while driving, the level of refinement can be set to maximum. But in other scenarios, such as changing a diaper, the refinement level can be lowered, transforming the words into a gentler version while preserving their meaning. This could allow caregivers to understand the intent behind verbal agitation and respond to any discomfort their PwD is experiencing.

\subsubsection{Prompting just-in-time maneuvers (information support)}

Prompting just-in-time guidance during unpredictable verbal agitation could be useful by enhancing the resilience of informal caregivers while they feel overwhelmed and uncertain about handling sudden, unpredictable situations. Timely guidance for informal caregivers can be implemented using a variety of devices and modalities, such as audio prompts from earbuds, text notifcations on a smartphone, visual guidance on AR devices, and haptic signals from smart watches. It can help them stay alert and take appropriate actions promptly.

\ipstart{Potential usage scenario}
Danielle is the primary caregiver for her mother with mixed dementia (a combination of Alzheimer's disease and vascular dementia). One day, her family takes her mother to a fine-dining restaurant to celebrate her mother's 81st birthday. Unexpectedly, her mother begins to curse at other patrons in the restaurant without any apparent reason. She makes derogatory and racist remarks, causing Danielle to feel extremely embarrassed and panicked, unsure of how to handle the situation. While being aware that she has to immediately stop her mother and apologized to people in the restaurant, Danielle is flustered and paralyzed by the sudden verbal agitation. A moment later, her smartwatch recognizes the surge of her stress level. Simultaneously, her smart glasses detects the context and identified the affected parties in the restaurant, then it presents visual prompts suggesting an immediate apology to the offended patrons (\textbf{information support}).

\ipstart{Design consideration}
Context-aware through smart glasses may bring privacy concerns. Alternatively, designers can consider using a microphone to capture contextual information with preserving privacy~\cite{iravantchi2021:privacy}. Also, informal caregivers may have different preferences in receiving information. Providing information in various cues at the same time can further overwhelm them particularly when they are panicked. Therefore, caregivers should be able to customize the form and modality of guidance prompts during such unforeseen incidents.

\subsection{Caring for Caregivers after Verbal Agitation}

After verbal agitation, informal caregivers strive to relieve their emotional stress (\autoref{tech:end1}) and to look for learning materials for future preparation (\autoref{tech:end2}). Informal caregivers may experience hurdles when doing so. First, informal caregivers' friends and families are not readily available for emotional support; when available, they may fail to provide appropriate emotional support. In response, we propose technology that can \textbf{comfort} informal caregivers to address their emotional stress. Second, descriptive learning materials online are less of help. We suggest technology that enables informal caregivers to \textbf{reflect} on their practices for learning purposes.

\subsubsection{Comforting informal caregivers to address emotional stress (social support; proximal)}

Technologies designed to comfort informal caregivers can help manage emotional stress and improve their mental health proximal after verbal agitation. Such comfort can be provided through various means such as conversation~\cite{baecker2020:emotion,kim2019:lyrics}, music~\cite{amy2004:music,guan2021:cue}, activities~\cite{ferguson2014:craving}, or photographs~\cite{gaver2011:photo}, as documented in multiple studies. These technologies can play a crucial role in reducing \mbox{emotional stress.}

\ipstart{Potential usage scenario}
Emily has been caring for her mother, diagnosed with Alzheimer's disease, for about eight years. Emily takes her mother to the hospital for a regular checkup. After completing their visit, her mother suddenly accuses a nurse of stealing her purse. However, she actually did not bring it from home. Emily feels embarrassed and apologetic towards the nurse as her mother yells at the hospital staff like: ``\textit{give me back my purse, you b**ch!}''. Once they return home, Emily seeks emotional support from her sister via a phone call, but unfortunately, she does not get a response. In her need for an outlet, she turns to Siri, a virtual assistant, sharing the details of the distressing hospital incident. Siri empathizes with Emily~\cite{jutten2019:empathy}, acknowledging the awkwardness of the situation and asking about her response~\cite{siemens2011:families} --- ``\textit{That sounds terrible. You must be embarrassed. How did you handle it?}'' It then suggests a song by Emily's favorite artist. This interaction leaves Emily feeling relieved (\textbf{social support}).

\ipstart{Design consideration}
Informal caregivers might find it uncomfortable to receive emotional support from a virtual agent for various reasons. Some people feel voice assistants creepy~\cite{rachel2023:creepiness}. One possible reason could be the agent's formlessness, which can feel unsettling. To address this, designers can integrate these conversational features into physical artifacts (\eg Paro~\cite{marti2006:socially}, Nao~\cite{shamsuddin2011:nao}). Furthermore, The voice of the virtual agent can affect how users perceive its creepiness. To address this, researchers might use the voices of the caregiver's family members~\cite{sam2021:kinvoices, slepian2018:personality}.

\subsubsection{Education through learning and practice (information support; distal)}

Technology aimed at educating informal caregivers could be valuable for improving informal caregivers' well-being distal after verbal agitation. This could involve teaching about dementia symptoms and suitable care methods. Initially, learning can occur through reflection~\cite{colin2013:enhancing}, although this can be challenging without records. Therefore, technology can facilitate opportunities for reflection. Such reflection could be made by recording the situation using mobile phones, smart earbuds, or virtual agents in any smart home device. Moreover, practice, which many caregivers find challenging (\eg P8), can also be supported by technology, serving as a simulated partner for rehearsals. The simulation practice could be made through a personal computer, a smartphone, a tablet screen, VR devices~\cite{lungu2021:vr}, or a virtual agent in smart home devices.

\ipstart{Potential usage scenario}
Felix lives with his father, who was diagnosed with senile dementia 12 years ago. As dementia is progressive, his father has started exhibiting verbal agitation since January. Felix is confused as he never experienced verbal agitation before. Felix attempts to learn more about dementia caregiving through Google and ChatGPT, but he finds applying the text-based knowledge he gains from the internet to real-time situations quite challenging. Felix discovers an AI-powered service that offers feedback and role-playing to help caregivers reflect on and practice handling verbal agitation. He uses his smartphone to record three typical situations involving verbal agitation. The service then analyzes these recordings and offers feedback, alongside mock simulations for practical application (\textbf{information support}). This allows Felix to implement what he learns from the feedback in a safe, controlled environment. This improved learning experience helped Felix gain confidence in handling verbal agitation.

\ipstart{Design consideration}
Designers must be aware that using AI technologies often brings up security and ethical concerns, which require further research. For instance, informal caregivers might worry about the handling of their voice data. Additionally, determining how to appropriately obtain, interpret, and apply the consent of informal caregivers and PwD's consent to share their data to improve learning materials for other informal caregivers or personalization for them would be another interesting future research question~\cite{jan2007:participatory,lena2023:working}.

\section{discussion and future work}

\subsection{Expanding the Design Space for Improving Informal Caregivers' Well-being}
Our study expands the boundaries of dementia-related research in the HCI community. Prior dementia-related studies mostly focus on exploring problems that PwD experience (\eg mobile phone use~\cite{dixon2022:phone}, information accessibility~\cite{dixon2022:barriers}, supporting daily activities~\cite{wang2017:daily}), which is crucial to improving the well-being of PwD. By improving PwD's well-being, we may expect indirect positive impacts on informal caregivers' well-being. However, it is also important to explicitly design solutions for problems that informal caregivers experience, such as verbal agitation from PwD. Dementia is progressive and brings negative impacts on the well-being of both informal caregivers and PwD. As dementia advances, many behavioral and psychological symptoms become severe, resulting in informal caregivers' well-being deteriorating. We found this design space holds lots of challenges to consider such as constraints on modalities (\autoref{dis:distraction}) and complicated social interactions (\autoref{dis:para}). Design opportunities in this paper provide the roadmap for future work to improve the well-being of informal caregivers experiencing verbal agitation from PwD.

\subsection{Reflections on Participants' Cultural Disparities}

In our study, while we initially aimed to collect inclusive data rather than make direct comparisons (\autoref{subsec:participants}), we discovered a few notable differences between participants from South Korea and the USA. For example, participants from South Korea showed resigned attitudes towards verbal agitation of PwD while those from the USA presented proactive attitudes. Korean participants (\eg~P3, P6) often mentioned how they endured verbal agitation of PwD, not necessarily seeking help or exploring alternative strategies. Instead, they seemed inclined to cope with the resulting mental distress after verbal agitation happened. On the other hand, participants from the USA (\eg~P7, P11) tended to learn and take maneuvers to resolve instances of verbal agitation and implement preventive strategies for verbal agitation.

This observation highlights the cultural disparities in attitudes and responses towards verbal agitation of PwD between Korea and the USA. Yet, we do not suggest generalizing these findings, as this study was not designed to investigate cultural differences. Instead, we underscore that informal caregivers' attitudes can vary. Thus, future work should take into account such possible variations and dimensions when designing technology for informal caregivers of PwD, ideally with further cultural domains (\eg~ethnicity, generation).

\subsection{Limitations and Future Directions}

Our study has several limitations. First, the recruitment process may have introduced participant bias. Participants from Korea were primarily recruited through word-of-mouth, while those from the USA were recruited from caregiver support communities in local libraries. Regular attendance at these support group meetings already implies persistence and a proactive approach to caregiving. Moreover, the ability to allocate time for these meetings may suggest a relatively more comfortable financial situation. It is worth noting that individuals with limited financial resources often face challenges balancing work commitments and caregiving responsibilities, making it difficult to participate in support communities. Hence, future research should consider diversifying the recruitment channels. 

Secondly, we only recruited our participants caring for PwD mostly having Alzheimer's disease (AD) among many other types of dementia. Since different behavioral and psychological symptoms depend on dementia types~\cite{chiu2006:symtomps}, the impact on the mental health of informal caregivers and their technological needs may vary. For example, although it is not common as Alzheimer's disease and vascular dementia, people with Lewy bodies dementia (LBD) also manifest verbal agitation~\cite{rebecca2001:dementia}. Considering people with LBD commonly exhibit hallucination in addition to verbal agitation~\cite{martine2000:review, liu2017:effects}, informal caregivers of people with LBD may perceive different burdens and need to address them in different ways. Therefore, future work may need to extend our findings by involving caregivers looking after PwD with different types of dementia. 

Finally, we only interviewed informal caregivers who may have limited expertise in the medical impact of such technologies on dementia symptoms and patient/caregiver relationships. Therefore, in a future study, we may conduct participatory research with multiple stakeholders, such as professional caregivers, medical doctors, and non-caregiver family members, to consider the possible impact of technology more broadly.

\section{conclusion}

We conducted interviews with 11 informal caregivers who suffer from verbal agitation of PwD. Our objective was to explore how technology could support informal caregivers' mental health. Through our interviews, we discovered that the impact of verbal agitation on informal caregivers varies depending on its predictability. Subsequently, we identified the coping mechanisms currently employed by informal caregivers, and their requirements in designing mental health support tools. We categorized these needs into time axes: before, during, and after verbal agitation. Based on these findings, we outlined potential design opportunities to enhance the mental well-being of informal caregivers suffering from verbal agitation. \textit{Before} impending verbal agitation, we proposed technology facilitating informal caregivers' preparation by offering either distractive stimuli (in-situ support) or encouragement (social \& information support). \textit{During} unpredictable verbal agitation, we suggested technology bolstering informal caregivers' resilience by alleviating the impact of verbal agitation (in-situ support) or prompting just-in-time maneuvers (information support). \textit{After} verbal agitation, we described technology that can comfort informal caregivers (social support) and help informal caregivers reflect on learning better care practices for future preparation (information support). Considering the outcomes of our research and discussions, we emphasize the need for further research in the HCI and CSCW communities to build artifacts that could improve the overall well-being of informal caregivers.


\begin{acks}
We deeply appreciate our participants for sharing their time and stories. We thank the anonymous reviewers for their constructive feedback. We also thank Duri Long and Darren Gergle for their valuable comments during the revision process. We acknowledge the early-stage feedback from John Joon Young Chung and Sooyeon Jeong.
\end{acks}

\bibliographystyle{ACM-Reference-Format}
\bibliography{bib}


\end{document}